	\documentclass[a4paper,fleqn,usenatbib]{mnras}
	
	\usepackage{txfonts}
	\usepackage[T1]{fontenc}
	\usepackage{ae,aecompl}
	\usepackage[center]{caption}
	\usepackage{graphicx}	% Including figure files
	\usepackage{float}
	\usepackage{hyperref}
	\usepackage[flushleft]{threeparttable}
	\title[AstroSat and MAXI view of Cygnus X-1]{{\it AstroSat} and {\it MAXI}
	view of Cygnus X-1: Signature of an `extreme' soft nature}
	
	\author[Ankur Kushwaha et al.]{
	Ankur Kushwaha$^{1,\,2}$\thanks{E-mail: ankurksh@ursc.gov.in},
	V. K. Agrawal$^{1}$%\thanks{E-mail: vivekag@ursc.gov.in}
	, Anuj Nandi$^{1}$%\thanks{E-mail: anuj@ursc.gov.in}
	\\
	$^{1}$Space Astronomy Group, ISITE Campus, U. R. Rao Satellite Centre,
	Outer Ring Road, Marathahalli, Bangalore, 560037, India\\
	$^{2}$Department of Physics, Indian Institute of Science, Bangalore, 560012, India\\
	}
	
	\date{Accepted XXX. Received YYY; in original form ZZZ}
	
	\pubyear{2020}
	
\begin{document}
	\label{firstpage}
	\pagerange{\pageref{firstpage}--\pageref{lastpage}}
	\maketitle
	
	% Abstract of the paper
	\begin{abstract}
	We present a detailed spectral and timing analysis of Cygnus X-1 
	with multi-epoch observations, during $2016$ to $2019$, by  {\it SXT}
	and {\it LAXPC} on-board {\it AstroSat}.
	We model the spectra in broad energy range of $0.5\!-\!70.0\,\rm{keV}$ 
	to study the evolution of spectral properties while Cygnus X-1 
	transited from hard state 
	to an extreme soft state via intermediate states in 2017.
	Simultaneous timing features are also examined by
	modelling the power density spectra in $3.0\!-\!50.0\,\rm{keV}$ .
	We find that during high-soft state observations, made 
	by {\it AstroSat} on Oct $24,\,2017$ (MJD $58050$), the energy 
	spectrum of the source exhibits an inner disk 
	temperature (kT$\rm_{in}$) of $0.46\!\pm\!0.01\,\rm{keV}$ ,
	a very steep photon index ($\Gamma$) of $3.15\!\pm\!0.03$
	along with a fractional disk flux contribution of $\sim\!45\%$.  	
	The power density spectrum in the range of $0.006\!-\!50.0\,\rm{Hz}$ 
	is also very steep with a power-law index of 
	$1.12\!\pm\!0.04$ along with a high RMS value of $\sim\!25\%$.   
	Comparing the spectral softness of high-soft state
	with those of previously reported,
	we confirm that {\it AstroSat} observed Cygnus X-1 in the `softest'
	state. The lowest {\it MAXI} spectral hardness
	ratio of $\sim\!0.229$ corroborates the softest nature of the source.
	Moreover, we estimate the spin of the black hole
	by continuum-fitting method, which indicates that Cygnus X-1 
	is a maximally rotating `hole'.
	Further, Monte Carlo (MC) simulations are performed to 
	estimate the uncertainty in spin parameter, which is 
	constrained as a$_{\ast}>0.9981$ with $3\sigma$ confidence 
	interval. Finally, we discuss the implications of our findings. 
	\end{abstract}

	
	%%%%%%%%%%%%%%%%%%%%%%%%%%%%%%%%%%%%%%%%%%%%%%%%%%
	
	%%%%%%%%%%%%%%%%% BODY OF PAPER %%%%%%%%%%%%%%%%%%
	
	\section{Introduction}
	\label{sec:intro}
	Black hole X-ray binaries (BH-XRBs) arguably provide wonderful 
	opportunities to understand accretion dynamics.
	In BH-XRBs, matter from companion star, forms an accretion 
	disk around a black hole. Matter loses angular momentum within 
	disk to viscous dissipation which causes inward drift, heating 
	and emission of radiation. Radiation emanating from disk in 
	proximity to black hole has signatures of strong gravity. 
	
	Emission from BH-XRBs can be of thermal and non-thermal origins. 
	The Keplerian accretion disk \citep{1973A&A....24..337S} 
	is considered to produce multi-colour thermal X-ray emission. 
	Inverse-comptonisation by a `hot' corona  
	\citep{1994ApJ...434..570T,1995xrbi.nasa..126T,1995ApJ...455..623C} 
	of 	soft photons emanating from disc, is responsible for 
	higher energy thermal or non-thermal emission.
	These two types of emissions are the major components 
	in the energy spectra, of which one component may dominate in a 
	spectral state of BH-XRBs \citep{2006ARA&A..44...49R}.
	
	The persistently bright BH-XRBs, in general, exhibit two 
	spectral states, high-soft state and low-hard state whereas 
	spectral states of outbursting BH-XRBs, based on hardness 
	intensity diagram (HID) and spectro-temporal features, can be 
	classified into low-hard state (LHS), hard-intermediate state (HIMS), 
	soft-intermediate state (SIMS) and high-soft state (HSS)
	\citep[and references therein]{2001ApJS..132..377H,
	2005Ap&SS.300..107H,2006ARA&A..44...49R,
	2012A&A...542A..56N, 2018Ap&SS.363..189D,
	2019MNRAS.487..928S, 2020MNRAS.497.1197B}.
		
	In HSS, energy spectrum is dominated by thermal emission 
	from disk over Compton tail, while in LHS it is vice-versa
	\citep{2006ARA&A..44...49R}.
	The thermal emission in HSS is of great importance as it
	originates in the inner regions of the disk and the spectrum
	has imprints of inner-disk radius (R$\rm{_{in}}$). R${\rm_{in}}$ is 
	effectively the inner most stable circular orbit (ISCO)
	\citep{1995xrbi.nasa..126T,2014SSRv..183..295M}.
	The radius of ISCO (R$\rm{_{ISCO}}$) is an important parameter 
	which in units of gravitational radii 
	(R$\rm{_{ISCO}}$/r$\rm{_{g}}$; r$\rm{_{g}} \equiv$ GM/$c^{2}$) is a simple 
	function of black hole spin 
	{\bf a$_{\ast}$}\footnote{\label{note:spin_starred}{\bf a$_{\ast}$}$\equiv$ 
	Jc/GM$\rm{_{BH}}^{2}$, $\mid${\bf a$_{\ast}$}$\mid$ $\leqslant$ 1, 
	where J is angular momentum	of black hole.}
	\citep{1972ApJ...178..347B}. Hence, by estimating R$\rm{_{in}}$, 
	black hole's spin parameter is measured in BH-XRBs.
	
	Mainly two methods have been developed to estimate the
	spin of black holes. First method is based on modelling
	of thermal continuum spectrum of black hole's accretion 
	disk \citep{1997ApJ...482L.155Z,2014SSRv..183..295M}
	and the second method estimates the spin by modelling 
	the profile of a relativistically broadened Fe-K$_{\alpha}$	
	fluorescence line \citep{1995Natur.375..659T, 
	2014SSRv..183..277R}. Alternatively, measurement of spin 
	is also possible via high frequency Quasi-periodic 
	Oscillations (HFQPOs) \citep[and references therein]{2001A&A...374L..19A,
	2006ARA&A..44...49R,2019MNRAS.484.3209D} and X-ray polarization
	\citep{2008MNRAS.391...32D,2009ApJ...701.1175S} 
	observed in BH-XRBs.
 
	Spin estimation	via X-ray continuum-fitting (CF) method
	requires a disk dominated HSS of BH-XRBs. The thermal 
	disk spectrum, is fitted with thin relativistic disk model
	{\tt kerrbb} \citep{1973blho.conf..343N,2005ApJS..157..335L}
	and in turn estimates R$\rm{_{in}}$ of the accretion disk. 
	R$\rm{_{in}}$, is then tagged as R$\rm{_{ISCO}}$ since it is widely accepted 
	that in this state the accretion rate is high enough for the disk
	to extend to R$\rm{_{ISCO}}$, which is related to spin  
	{\bf a$_{\ast}$}\textsuperscript{\ref{note:spin_starred}}.	
	Accurate measurements of distance (D), mass of black hole 
	(M$\rm{_{BH}}$) and inclination angle of the binary plane ({\it i}) 
	are crucial in order to determine black hole spin with CF method. 
	In order to reliably estimate the spin with CF method,
	a geometrically thin disk and a weak Comptonisation of thermal 
	seed photons is preferred. The disk luminosity satisfying a 
	criteria of $L/L_{Edd}<30\%$ indicates presence of
	thin disk \citep{2006ApJ...652..518M, 2014SSRv..183..295M}.
	The strength of Comptonising medium is 
	estimated with scattering fraction ({\it f$_{sc}$}), which indicates 
	the fraction of thermal seed photons that is 
	scattered to produce high energy tail. The condition
	{\it f$_{sc}$}$<25\%$ is necessary for successful
	application of CF method \citep{2009ApJ...701L..83S}.		
	The method has been employed for many BH-XRBs previously	
	\citep[and references therein]{2009ApJ...701.1076G,
	2014SSRv..183..295M} including Cygnus X-1
	\citep{2011ApJ...742...85G,	2014ApJ...790...29G, 
	2017PASJ...69...36K}, to measure the spin of the black hole.	
	
	Cygnus X-1 is a persistent bright high mass X-ray binary 
	with a primary object as confirmed black hole 
	\citep{1972Natur.235...37W, 1972NPhS..240..124B} and	
	a massive companion of 09.7 Iab type supergiant star 
	HDE $226868$ \citep{1973ApJ...179L.123W}. Previously, the binary bf was
	estimated to be at distance of 
	$1.86^{+0.12}_{-0.11}$ kpc based on radio parallax 
	\citep{2011ApJ...742...83R} and at $1.81\pm0.09$ kpc 
	based on dust scattering method \citep{2011ApJ...738...78X}.
 	The mass of the black hole in the binary system was estimated as 
 	M$\rm{_{BH}}$ $= 14.8\pm1.0$ M$_{\sun}$ and the inclination of 
 	binary plane as {\it i} $=27.1\pm0.8$ 
 	\citep{2011ApJ...742...84O}. Recently,  \citet{2021Sci...371.1046M},
 	using VLBA observations,  re-estimated the binary system 
 	parameters, M$\rm{_{BH}}$,
 	{\it i} and D to be $21.2\pm2.2,  27.51^{+0.77}_{-0.57}$ and $2.22^{+0.18}_{-0.17}$,
 	respectively.
 	Based on the previous parameters, the spin of the black hole,  was 
 	estimated 	by \citet{2014ApJ...780...78T} as $0.9882\pm0.0009$
 	($90\% $ confidence level) with Fe-K$_{\alpha}$ line fitting
 	method using {\it Suzaku } and {\it NuSTAR} data. 
 	\citet{2014ApJ...790...29G} employed CF method and 
 	estimated spin as a$_{*}> 0.983 $ at a confidence level of 
 	$3\sigma$ $(99.7\%)$.
 	\citet{2017PASJ...69...36K} determined the spin 
 	$0.95\pm0.01$ with CF method using data from {\it Suzaku}.
 	Recently, \citet{2021ApJ...908..117Z}, used the updated parameters
 	of binary system and revised the spin parameter as 
 	a$_{*}> 0.9985 $ at a confidence level of $3\sigma$.

	Cygnus X-1 mostly remains in LHS but displays transitions
	to HSS. HSS duration can be few months to years
	\citep{2014A&A...565A...1G}. 
	The LHS spectra can be modelled well with Comptonised 
	continuum model. High resolution broad-band spectroscopic 
	observations revealed an additional component attributing 
	to reflection of Comptonised photons from disk 
	\citep{1983ApJ...275..307L, 2015ApJ...808....9P}.
	\citet{2017MNRAS.472.4220B} argued for presence of 
	an in-homogeneous Comptonising cloud and suggested two 
	Comptonising components for modelling the spectra. Spectra 
	in HSS, exhibit multi-colour blackbody accountable for 
	accretion disk with steep weak Compton tail. An additional 
	reflection component have also been reported in this state 
	\citep{1999MNRAS.309..496G,2014ApJ...780...78T, 
 	2016ApJ...826...87W,2017PASJ...69...36K}
 	of the source. Moreover, the broadband spectra
 	of both states have been modelled with two component 
 	advective flows model as well \citet{2007Ap&SS.309..305M}.
 	 	
	The short term and long term temporal features and their 
	evolution in different states of Cygnus X-1 have been 
	reported mainly from high time resolution {\it RXTE}
	and {\it AstroSat} observations
 	\citep{2003A&A...407.1039P,2005A&A...438..999A, 
 	2014A&A...565A...1G,2017ApJ...835..195M}.
 	The power density spectra (PDS) exhibit multiple broad 
 	features which are modelled 
 	with multiple Lorentzians. Exception to this are the 
 	`canonical' soft states when one lorentzian plus power-law or 
 	only a power-law can successfully represent the PDS 
 	\citep{2001MNRAS.321..759C,2005A&A...438..999A}.
 	During transitions from LHS to HSS, the central	
 	frequencies of Lorentzians shift to higher frequencies
 	\citep{2005A&A...438..999A} but remain below $25$ Hz.
 	
	 \begin{figure*}
	 	\includegraphics[width=\textwidth]{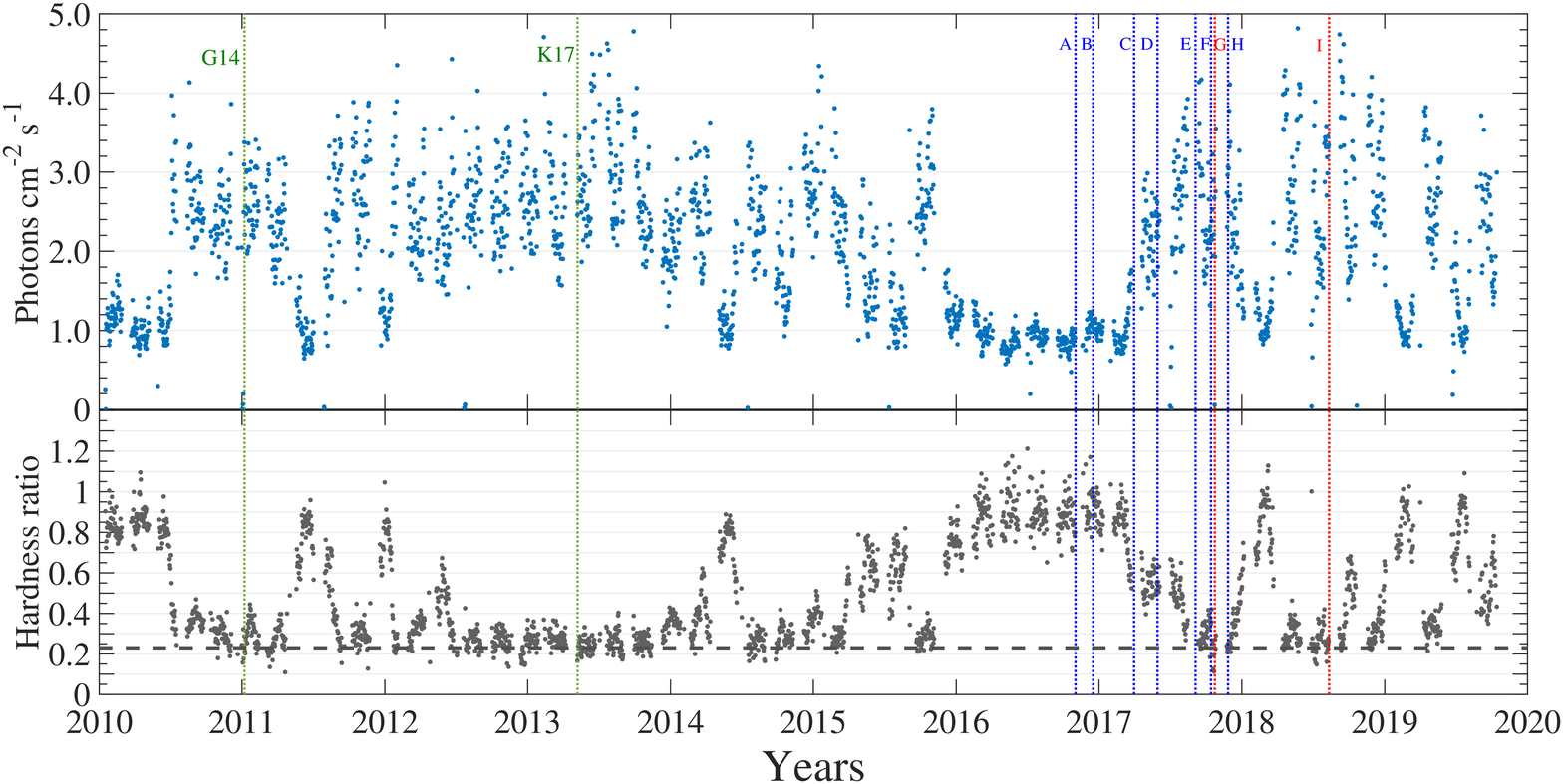}
	    \caption{The lightcurve of Cygnus X-1 in the $2.0-10.0$ keV band (top) 
	    and hardness ratio (bottom) based on data obtained using
		the {\it MAXI} Gas Slit Camera (GSC; \citet{2011PASJ...63S.623M}).
		The hardness ratio is defined as the ratio of counts 
		detected in a hard X-ray band ($4.0-10.0$ keV) to those 
		detected in a soft band ($2.0-4.0$ keV). The lightcurve and hardness
		ratio are day averaged. Green dotted lines 
		denoted with G14 and K17 show previous measurements of spin by
		\citet{2014ApJ...790...29G} and \citet{2017PASJ...69...36K}. 
		Blue and red dotted lines show {\it AstroSat} observations 
		along with alphabets corresponding to \autoref{tab:obs_table},
		used for this work. Red lines mark the epochs, considered for 
		spin measurement of Cygnus X-1 (see $\S$\ref{sec:softest_spec_analysis}).
		Horizontal black dashed line marks the lowest
		hardness ratio of $0.229$
		during {\it AstroSat} Obs G.}
	\label{fig:maxi_plot_figure}	
	\end{figure*}
	
	\begin{table*}
			\caption{{\it AstroSat} observations of Cygnus X-1 considered in present analysis.
			From left to right,
			(1) Name of observation; 
			(2) corresponding {\it AstroSat} observation ID;
			(3) spacecraft orbit number since launch date;
			(4) date of observation;
			(5) {\it LAXPC} exposure time;
			(6) {\it LAXPC}	 count rate;
			(7) {\it SXT} exposure time;
			(8) {\it SXT}	 count rate. 
			{\it LAXPC} and {\it SXT} count rates, provided here, are post data reduction and 
			averaged over exposure time. See $\S$\ref{sec:obs_data} for details.}
			\label{tab:obs_table}
			\begin{tabular}{lcccccccccccc}
			\hline
			\hline
			Observation  & Obs ID    & Orbit      & Obs date (MJD)  & \multicolumn{2}{c}{LAXPC}& \multicolumn{2}{c}{SXT}\\
			&&&&(ksec)&(counts/sec)&(ksec)&(counts/sec)\\  
			\hline
			Obs A &   $9000000768$ &     $5931$   &  $2016-11-01\,(57693)$   
			&   $\sim3.36$  &$\sim1134$&$\sim2.34$&$\sim26$\\
			Obs B &   $9000000890$ &     $6597$   &  $2016-12-16\,(57738)$   
			&   $\sim2.77$  &$\sim1567$&$\sim0.9$&$\sim41$\\
			Obs C &   $9000001122$ &     $8152$   &  $2017-03-31\,(57843)$   
			&   $\sim4.33$  &$\sim1470$&$\sim1.87$&$\sim122$\\
			Obs D &   $9000001258$ &     $9043$   &  $2017-05-31\,(57904)$   
			&   $\sim3.73$  &$\sim1447$&$\sim1.46$&$\sim118$\\
			Obs E &   $9000001516$ &    $10485$   &  $2017-09-05\,(58001)$   
			&   $\sim5.10$  &$\sim1389$&$\sim2.14$&$\sim263$\\
			Obs F &   $9000001616$ &    $11068$   &  $2017-10-14\,(58040)$   
			&   $\sim2.86$  &$\sim1272$&$\sim1.57$&$\sim388$\\
			Obs G &   $9000001636$ &    $11212, 11215$   &  $2017-10-24\,(58050)$   
			&   $\sim3.19$  &$\sim990$&$\sim1.93$&$\sim381$\\
			Obs H &   $9000001726$ &    $11717$   &  $2017-11-27\,(58084)$   
			&   $\sim4.93$  &$\sim1938$&$\sim2.33$&$\sim402$\\ 
			Obs I &   $9000002302$ &    $15548$   &  $2018-08-13\,(58344)$   
			&   $\sim4.88$  &$\sim1162$&$\sim2.15$&$\sim291$\\
			\hline
			\end{tabular}
	\end{table*}
		
	Cygnus X-1 has been observed multiple times over last $5$ 
	years with {\it AstroSat}, India's first multi-wavelength space 
	observatory \citep{2006AdSpR..38.2989A}. {\it AstroSat} was 
	launched on Sept 25, $2015$, with five scientific instruments 
	on-board covering a wide range of energies from UV to hard X-rays 
	\citep{2014SPIE.9144E..1SS, 2016arXiv160806051R}.
	It provides a unique platform for simultaneous observations, 
	over a wide X-ray band from $0.3$ keV to $100$ keV, via its suite 
	of co-aligned X-ray instruments - Soft X-ray Telescope ({\it SXT})
	 \citep{2016SPIE.9905E..1ES, 2017JApA...38...29S},
	Large Area X-ray Proportional Counter ({\it LAXPC}) 
	\citep{2016SPIE.9905E..1DY, 2017ApJS..231...10A} 
	and Cadmium Zinc Telluride Imager ({\it CZTI}) 
	\citep{2016SPIE.9905E..1GV}. 
	We exploit the detection capabilities of {\it SXT} in the soft X-rays 
	to obtain spectra in the lower energies and {\it LAXPC} enables us to 
	study spectra up to $80$ keV. We studied evolution of spectral 
	and temporal properties of Cygnus X-1 over a transition from LHS 
	to HSS via intermediate states (IMS). Cygnus X-1 attains
	an extreme soft state and we find that during this
	state source has the `softest nature' ever
	observed by any other X-ray observatory. Further, we employ CF 
	method to estimate the spin of the black hole. 
	Since accuracy and success of CF method rely on broadband X-ray 
	observation of BH-XRBs, {\it AstroSat} data is suitable for 
	such studies. We also make use of the temporal resolution
	($\sim10$ $\mu$s) of {\it LAXPC} to obtain PDS of Cygnus X-1,
	in order to study the temporal features during the 
	softest state and various phases of accretion states.
	
	In $\S$\ref{sec:obs_data}, we have given details of the observations 
	and the steps for data reduction. We have discussed broadband spectral 
	analysis using both {\it SXT} and {\it LAXPC} data and the results 
	from phenomenological as well as physical modelling in 
	$\S$\ref{sec:spectral_analysis}. 
	Detailed analysis of PDS has been carried out with 
	{\it LAXPC} data and the results are presented in 
	$\S$\ref{sec:timing}. We discuss about the `softest' nature of 
	the source and present the details of spin measurement in the 
	softest state with CF method in $\S$\ref{sec:softest_spec_analysis}.
	Finally, we have discussed the results and have concluded in 
	$\S$\ref{sec:discussion}.

	\section{Observations and Data Reduction}
	\label{sec:obs_data}
	
	Cygnus X-1, for the last decade $2010-2020$, has been showing 
	atypical behaviour of remaining mostly in soft states
	when compared to previous
	$\sim\!15$ years $(1996-2010)$ where it was typically found 
	in hard state with occasional soft state transitions 
	\citep{2014A&A...565A...1G}. Lightcurve obtained from continuous
	monitoring of source with {\it MAXI}, clearly 
	shows this behaviour of the source (see \autoref{fig:maxi_plot_figure}).
	The monitoring shows that the source is in persistent LHS 
	in $2016$ and subsequently, in $2017$ starts a transition into HSS.
	We make use of {\it MAXI} lightcurve in $2.0-10.0$ keV energy band  
	and hardness ratio (HR) of two energy bands $4.0-10.0$ keV to $2.0-4.0$ keV 
	(hereafter {\it MAXI}-HR) in order to select observations from 
	{\it AstroSat} data during this
	transition period (\autoref{fig:maxi_plot_figure}). 
	
	Cygnus X-1 was observed by {\it AstroSat} for last $5$
	years ($2016-2020$) with about $\sim60$ co-ordinated pointings. 
	We have selected $9$ observations from Guaranteed Time (GT) 
	phase for this work, as our primary goal is to study the broadband 
	characteristics during the transition phase of the source.
	Hence, we select only those observations which are suitable to 
	study the entire transition from LHS to HSS via IMS during the 
	AstroSat campaign. The details of multi-epoch 
	observation logs are provided in \autoref{tab:obs_table} and also marked in 
	\autoref{fig:maxi_plot_figure}. We make use of data from {\it SXT} and 
	{\it LAXPC} on board {\it AstroSat} to carry out broadband spectral and timing
	analysis on these selected observations.
	   	
\begin{figure}
		\hskip 3 mm
		\includegraphics[width=\columnwidth]{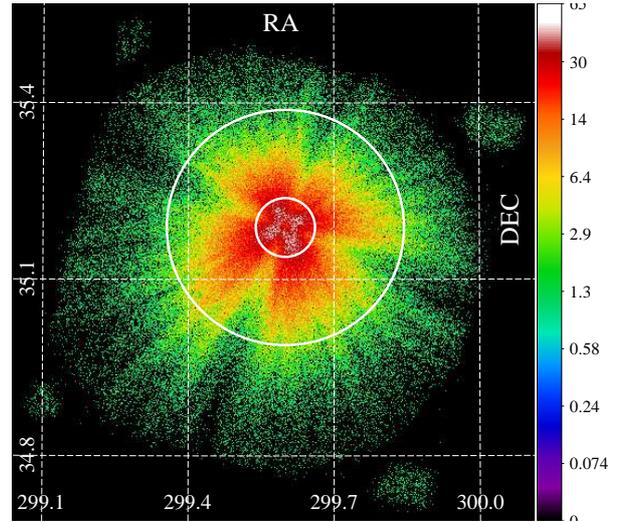}
	    \caption{The image of Cygnus X-1 on {\it SXT} CCD shown in false colors 
	    during Oct $24\,2017$ (MJD $58050$) observation. CCD was operated in photon
	    counting (PC) mode. Source photons are extracted within annular 
	    region, shown in solid white, with inner and outer radii
	    of $3\arcmin$  and $12\arcmin$, respectively. The region within 
	    inner radius is excluded to avoid piled-up events. Calibration 
	    source illumination can be seen on four corners of the CCD.
	     The colour bar alongside is for the source intensity in {\it SXT} 
	     (cts pixel$^{-1}$). See text for details.}
	    \label{fig:sxt_image}
	\end{figure}

   	{\it SXT} is a soft X-ray instrument capable of imaging and spectroscopy 
   	in the $0.3-8.0$ keV energy range. {\it SXT} has a focusing telescope 
   	and a charged-coupled device (CCD) detector \citep{2017JApA...38...29S}.
   	The SXT CCD calibration procedures and results are summarized by
   	\citet{2016SPIE.9905E..1ES, 2017JApA...38...29S}.
   	Extraction of level $2$ data from {\it SXT} raw data is done using the 
   	{\it sxtpipeline} tool by {\it SXT} instrument 
   	team\footnote{\label{note:sxt_weblink}\url{https://www.tifr.res.in/astrosat_sxt/dataanalysis.html}}
   	and is available on ISSDC 
   	website\footnote{\label{note:issdc_weblink}
   	\url{https://www.issdc.gov.in/astro.html}}
   	\citep[see also][]{2019MNRAS.487..928S, 2020MNRAS.497.1197B}.
   	The {\it SXT} data are available in the Photon Counting (PC) mode. 
   	The extracted cleaned event files are used to generate energy spectra 
   	with {\it XSELECT V2.4d}. Spectra and lightcurve are extracted from 
   	an annular region over {\it SXT} CCD image of the 
   	source (see \autoref{fig:sxt_image}). As Cygnus X-1 is a 
   	bright source and saturates the central pixels of image location on 
   	{\it SXT} CCD, thus annular source region with inner radius of 
	$\sim 3\arcmin$ has been excluded for data extraction to avoid pile-up	
	(see $\S$\ref{sec:discussion}, for details on pile-up estimation
	in SXT spectra),
	while the outer radius varied from $12\arcmin-16\arcmin$ depending 
	on the size of source image in different states and epochs.
	Smaller outer radii correspond to LHS data sets since counts from
	source is very low and bigger radii causes high background contribution.
	The spectra have 1024 channels and for fitting purpose, we use
	the spectra without rebinning along with the background, 
	response as well as ARF files provided 
	by {\it SXT} instrument team\textsuperscript{\ref{note:sxt_weblink}}.

   The {\it LAXPC} is one of the primary instruments on board {\it AstroSat} 
   and consists of three identical co-aligned X-ray proportional counter 
   units ({\it LAXPC10}, {\it LAXPC20} and {\it LAXPC30}) providing with high
   time resolution ($\sim10$ $\mu$s) covering $3.0-80.0$ keV energy band 
   \citep{2016SPIE.9905E..1DY, 
   2017ApJS..231...10A, 2017JApA...38...30A}. All three units are calibrated
   with Crab observations and further details on the same are given by 
   \citet{2017ApJS..231...10A,2021arXiv210107514A}.
   %The combined effective area of the three units is $\sim 6000$ cm$^{2}$ at 15 keV.
   Data from {\it LAXPC10} unit are unstable, while {\it LAXPC30} data are
   not considered due to the continuous gain shift observed in this unit, 
   suspected to be caused by a gas leakage \citep{2017ApJS..231...10A}. 
   Thus, in this work, we use the data only from {\it LAXPC20}, which
   is stable and working nominally since launch. A very recent Crab observation
   with {\it LAXPC20} is discussed by \citet{2021arXiv210107514A}
   where spectral fit parameters are found to be well within 
   the acceptable limits.  
   {\it LAXPC20} was operated in event analysis (EA) mode during all 
   the observations.
   The  capabilities of {\it LAXPC} for timing and spectroscopic studies 
   of various sources are already demonstrated by 
   \citet{2016ApJ...833...27Y, 2017ApJ...835..195M, 
   2017ApJ...841...41V, 2018MNRAS.477.5437A, 2019MNRAS.487..928S,
   2020MNRAS.497.1197B, 2020MNRAS.497.3726A, 2020Ap&SS.365...41A,
   2020arXiv201003782S}.
		    
   We followed these papers and the instructions provided along with the 
   {\it LAXPC} analysis software 
   ({\it LaxpcSoft}\footnote{\url{https://www.tifr.res.in/~astrosat_laxpc/software.html}})
   released on Sept $9,\,2019$. The details of how the responses and background 
   spectra are generated, are described in \citet{2017ApJS..231...10A}.
   Further, the source and background energy spectra are extracted 
   from events in all layers and anodes of {\it LAXPC20}. Moreover, 
   only single events are considered for generating the spectra.
   These options are selected to have a minimal undesired bump in the spectral
   residuals at around $33$ keV which attributes to Xenon fluorescence 
   from detector. The extracted spectra have 256 channels and we use
   the spectral data for fitting without rebinning.
   {\it LaxpcSoft} generates GTI files of good time 
   interval which reduces the data and results in data gaps attributing 
   to South Atlantic Anomaly (SAA) passes and Earth occultations
   during observations.
   
	\section{Spectral Analysis and Results}
	\label{sec:spectral_analysis}
	\begin{figure*}
		\includegraphics[scale=0.51]{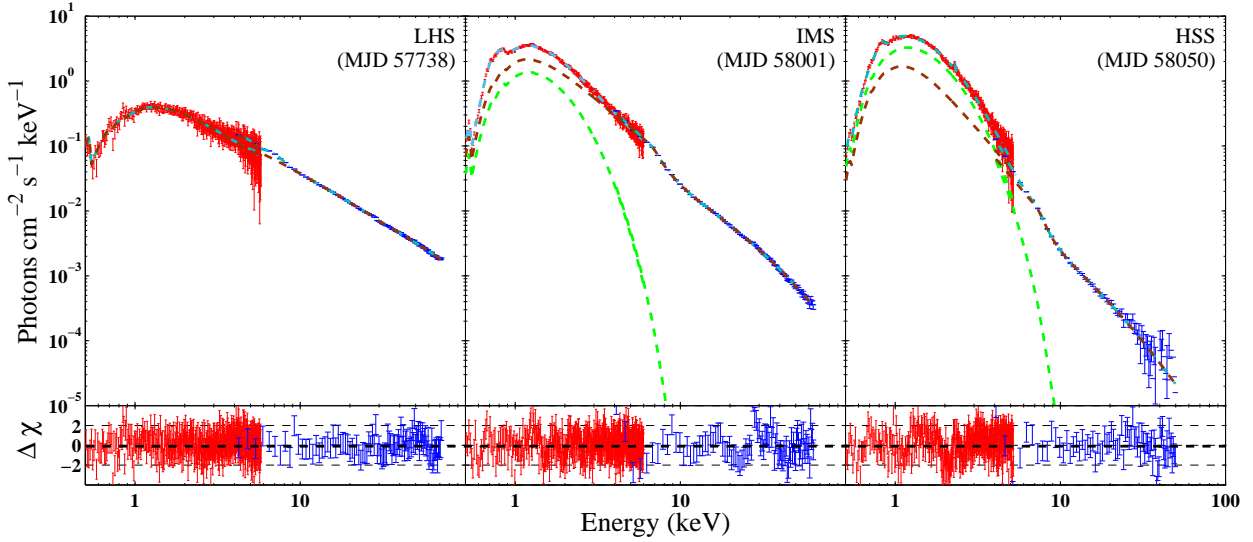}
	    \caption{Broadband energy spectra (unfolded) of Cygnus X-1 in LHS (left), 
	    IMS (middle) and HSS (right)  
	    for Obs B, E \& G, respectively. {\it SXT} and {\it LAXPC } data is
	    shown in red and blue, respectively. Hard state spectra are fitted 
	    with absorbed {\it nthComp} (dashed brown line). There is an additional 
	    {\it diskbb} component (green dashed line), for the other two states (IMS and HSS).
	    %Gaussian profile of Iron line is shown in dashed orange line. 
	    Cyan dotted dash lines show the total model. Bottom panels show the 
	    corresponding residuals in units of $\sigma$. See the text for details.}
	    \label{fig:spectra_figure}
	\end{figure*}

	We consider broadband modelling of energy spectra obtained during all 
	observations listed in \autoref{tab:obs_table}. {\it SXT} and 
	{\it LAXPC} spectra are simultaneously 
	fitted in {\it HEAsoft XSPEC 12.10.0e} package. Energy ranges of 
	$0.5-6.0$ keV and $4.0-70.0$ keV have been considered in spectral fit
	for {\it SXT} and {\it LAXPC}, respectively. 
	
	We start with Obs A and 
	modeled the spectrum with {\tt Tbabs} \citep{2000ApJ...542..914W}, 
	{\tt diskbb} \citep{1984PASJ...36..741M} and {\tt power-law}
	of {\it XSPEC}.
	The relative cross-normalization between the {\it LAXPC} and {\it SXT} 
	data is taken care by multiplicative constant factor. The fit gives a very 
	low and unrealistic value of disk temperature though $\chi_{red}^{2}$ 
	of $927.76/643=1.44$ indicates a good fit.
	Hence, we removed the disk component and fitted with {\tt Tbabs} and 
	{\tt nthComp} \citep{1996MNRAS.283..193Z}.
	The model gives a $\chi_{red}^{2}$ of $1077.5/645=1.67$. 
	The value of kT$\rm{_{bb}}$ is fixed to a fiducial value of $0.1$ keV.
	We then correct for instrument 
	features at $1.8$ and $2.24$ keV arising from {\it SXT}.
	Also, a broad absorption dip, present at $\sim 8$ keV,
	is modelled with {\tt smedge}. An instrumental Xenon
	{\tt edge} around $30$ keV and $2\%$ systematic uncertainty 
	is introduced in order to achieve a good estimate of the 
	spectral fit parameters \citep{2019MNRAS.487..928S}.  
	This results into a significantly improved $\chi_{red}^{2}$ 
	value of $848.72/638=1.33$. The fit gives model parameters of photon index 
	($\Gamma$) as $1.59\pm0.01$ and electron temperature (kT$_{e}$)
	of $86.94_{-21.91}^{+81.89.}$. Unless mentioned explicitly, all the 
	errors are computed using  $\Delta\chi^{2}=1.0$ ($1\sigma$ confidence level) 
	for all the observations. Spectral analysis and fit procedure for Obs B
	is similar to that of Obs A. The resulting fit parameters are
	given in \autoref{tab:fit_param_table} and the spectra fitted with the model
	is shown in the left of \autoref{fig:spectra_figure}.
	
	The hydrogen column density N$_{H}$ is estimated to be
    $7.52\times10^{21}$ atoms cm$^{-2}$ from the fit. This is in good 
    agreement with  estimations of previous investigators \citep{, 2014ApJ...780...78T, 
 	2015ApJ...808....9P, 2016ApJ...826...87W, 2017PASJ...69...36K}. It is to
 	be noted that HEASARC N$_{H}$ 
 	calculator\footnote{\label{note:nh_web_tool}\url{https://heasarc.gsfc.nasa.gov/cgi-bin/Tools/w3nh/w3nh.pl}}
    gives an average value for hydrogen column density of $7.01 \times 10^{21}$ 
    atoms cm$^{-2}$ for Cygnus X-1. \citet{2011ApJ...738...78X} estimates hydrogen 
    column density towards Cygnus X-1 as N$_{H}\approx4.65-4.85\times 10^{21}$ 
    atoms cm$^{-2}$ by dust scattering modelling. 
 	
	Moving onto the softer observations, as indicated by hardness ratio from
	{\it MAXI} (see \autoref{fig:maxi_plot_figure} and \autoref{tab:obs_table}),
	we attempt to fit the spectrum, 
	from Obs C, with {\tt Tbabs} and {\tt nthComp},
	but the fit shows residuals below 10 keV, suggesting inclusion of a thermal disk 
	emission component in the model. Hence, {\tt diskbb} is added to model the 
    spectrum. The fit gives a value of $0.29\pm0.01$
    for inner disk temperature (kT$\rm{_{in}}$)
    and a disk normalisation of $18.18^{+3.16}_{-2.76}\times10^{4}$. 
    The photon index ($\Gamma$) is found to be $1.91\pm0.01$, which suggests the spectrum 
    has become steeper. We note that Obs C marks the advent of
    soft state transition. Further, similar fit procedure is applied
    to Obs D and E.
    The inner disk temperature, during Obs D is slightly higher than that of Obs C and the 
    disk normalization decreases to a value of $12.26^{+2.52}_{-1.80}\times10^{4}$.
    Further, during Obs E, inner disk temperature increases to a value of $0.43\pm0.01$ while
    the fit gives a disk normalization of $2.94^{+0.28}_{-0.26}\times10^{4}$.
    Photon index ($\Gamma$) keeps on increasing, making the spectrum steeper 
    (see middle panel of \autoref{fig:spectra_figure}) with
    values of $1.86\pm0.01$ to $2.28\pm0.02$ for
    Obs D and Obs E, respectively. 
    Subsequently, spectral fit for Obs F 
    with the same model gives roughly similar values of inner disk temperature 
    and normalization of disk as Obs E. The photon index is even steeper with value 
    of $2.66\pm0.01$. 
    The spectrum from Obs G is fitted in the same way except the 
    energy range of {\it LAXPC} is considered from $5.0$ keV
    onwards. The fit results an even higher inner disk temperature
    of $0.46\pm0.01$ keV. Also, the fit results in  higher photon index
    ($\Gamma$) of $3.15\pm0.03$, which suggests the steepest spectrum of 
    all the observations (see right panel of \autoref{fig:spectra_figure}). 
    Fit results into a $\chi_{red}^{2}$ value of $946.08/530 = 1.78$. 
    Additionally, observations Obs H and I are modelled in 
    the same way as mentioned for previous states. 
    The fit parameters of all the observations are 
    summarized in \autoref{tab:fit_param_table}.
    
    \begin{table*}
    \begin{center}
    \begin{threeparttable}
			\caption{Spectral fit results with non-relativistic model 
			{\tt tbabs(diskbb+gaussian+nthComp)}\tnote{a}. From left to right, are
			(1) name of observation
	   		(2) hydrogen column density in units of $\times 10^{22}$ atoms cm$^{-2}$;
		    (3) inner disk temperature in units of keV; 
		    (4) normalization constant for {\tt diskbb} in units of $\times 10^{4}$;
		    (5) photon power-law index;
		    (6) electron temperature in units of keV; 
		    (7) seed photon temperature in units of keV;
		    (8) normalization constant for {\tt nthComp};
		    (9) unabsorbed flux, calculated in $0.1-100.0$ keV energy range, in units 
		    of $\,\times 10^{-8}$ ergs cm$^{-2}$ s$^{-1}$;
		    (10) $\chi_{red}^{2}\,(\chi^{2}$/dof; dof $\equiv$ degree of freedom) resulting from fit.
		    See text for details.}
		    
			\label{tab:fit_param_table}
	%		\begin{center}			
	%		\scalebox{0.95}{
	%		\begin{threeparttable}
			\begin{tabular}{lccccccccc} 
		    \hline
			\hline
			Obs &  N$_{H}$ &  kT$_{in}$  & N$_{disk}$	& $\Gamma$ 
			& kT$_{e}$ & kT$_{bb}$ & N$_{nthComp}$  & Flux & $\chi_{red}^{2}$($\chi^{2}$/dof)\\ 
			\hline
			Obs A       &  $0.75\pm0.01$	& - 		& - 		& $1.59\pm0.01$ 	
			& $86.94_{21.91}^{+81.89}$ & $0.1$ (fixed)	& $0.84\pm0.01$ &$2.56$ &$1.33(848.72/638)$\\
			Obs B       &  $0.54\pm0.01$  & -			& - 		&  $1.59\pm0.01$
			& $46.99_{-3.31}^{+1.71}$	& $0.1$ (fixed) & $1.15\pm0.01$ &$3.49$ & $1.18(739.52/624)$	\\
			Obs C       &  $0.73_{-0.02}^{+0.03}$	& $0.29\pm0.01$ 	& $18.18_{-2.76}^{+3.61}$ & $1.91\pm0.01$
			& $120$ (fixed)& $0.16_{-0.06}^{+0.12}$ & $2.64_{-1.02}^{+0.18}$ & $6.03$ & $1.01(627.13/620)$\\
			Obs D       &  $0.73_{-0.03}^{+0.04}$  & $0.31\pm0.01$  	& $12.66_{-1.80}^{+2.52}$ &	$1.86\pm0.01$	
			& $80$ (fixed) & $0.17_{-0.04}^{+0.20}$ & $2.22_{-0.70}^{+0.14}$ & $6.23$ & $1.19(750.44/627)$\\
			Obs E       &  $0.67\pm0.01$ & $0.43\pm0.01$ & $2.94_{-0.26}^{+0.28}$ & $2.28\pm0.01$
			& $43.99_{-7.76}^{+13.27}$ & $0.16\pm0.01$ & $7.33_{-0.33}^{+0.30}$ & $6.35$ &$1.48(927.65/624)$\\
			Obs F       &  $0.7$ (fixed)  	& $0.43\pm0.01$	& $3.26_{-0.19}^{+0.21}$	&	$2.66\pm0.01$
			&	$300$ (fixed) & $0.13\pm0.01$ &$15.46_{-0.42}^{+0.34}$ &$7.55$ & $1.56(990.00/633)$\\
			{\bf Obs G }&  $0.69\pm0.01$ & $0.46\pm0.01$ & $3.45\pm0.20$ &	$3.15\pm0.03$	
			& 	$300$ (fixed) & $0.15\pm0.01$ & $10.92_{-0.58}^{+0.60}$ & $7.17$ & $1.78(946.08/530)$\\
			Obs H       &  $0.7$ (fixed)	& $0.40\pm0.01$ & $4.94_{-0.27}^{+0.28}$	& $2.64\pm0.01$
			&	$150$ (fixed) & $0.13\pm0.01$ & $15.34_{-0.25}^{+0.21}$ &$9.15$ & $1.67(1021.92/611)$\\ 
			Obs I       	&  $0.69\pm0.01$& $0.45\pm0.01$ & $4.43_{-0.15}^{+0.13}$& $2.57\pm0.01$
			&	$300$ (fixed) & $0.13\pm0.01$ & $4.95_{-0.20}^{+0.16}$ & $6.26$ & $1.80(1072.92/594)$\\ 
			\hline
			\end{tabular}	
					
			\begin{tablenotes}
	   		\item[a] Note: The model is multiplied with {\tt smedge}  
	   		to represent the absorption feature. The component consists of
	   		a threshold energy ({\it E$_{c}$}), maximum absorption factor at 
	   		threshold ({\it f}), index for photo-electric 
	   		cross-section ($\alpha$) and smearing width ({\it w}). The parameters 
	   		{\it E$_{c}$}, {\it f}, $\alpha$ and {\it w} are found to 
	   		be varying in range of 7.3-8.0 keV, 0.5 to 3.6, -1.3 to -2.9 and 2.4 to 9.6 keV,
	   		respectively.
	   		%Also, {\tt edge} component is multiplied to model the Xe edge
	   		%feature arising from the instrument, as {\it LAXPC} instruments are large Xe 
	   		%filled gas proportional counters. The threshold energy of the component 
	   		%is at Xe edge i.e. $\sim$ 30 keV.}
	   		
			\end{tablenotes}
		    \end{threeparttable}

			\end{center}
		\end{table*}
    
    \subsection{The Softest Spectral State  }
    \label{sec:softest_state}
    
    The spectral fit parameters 
    obtained in the analysis evidently confirm that the spectral
    nature of the source evolved from LHS to HSS. Additionally, 
    in order to identify the spectral states of the source, we define
    criteria based on the inner disk temperature (kT$\rm{_{in}}$) 
    and photon index ($\Gamma$). We classify observations with
    $\Gamma< 1.8$ and no disk component, as LHS. Spectra that exhibit
    $\Gamma$ between $1.8-2.5$ along with kT$\rm{_{in}}$ in  range
    $0.3-0.4$ keV are identified as IMS. The observations 
    with very steep $\Gamma$ values in $2.5-3.2$ and high kT$\rm{_{in}}$
    ranging in $0.4-0.5$ keV are classified as HSS. Hence, Obs A \& B 
    are classified as LHS, Obs C, D \& E as IMS and Obs F, G, H \& I are
    of HSS.
    
    Moreover, the spectral fit parameters 
    obtained for observation made by {\it AstroSat} on
    Oct $24,\,2017$ (Obs G, MJD $58050$) clearly indicate that 
    Cygnus X-1 exhibits very soft nature as the spectrum 
    has the highest inner disk temperature of $0.46\pm0.01$
    and the steepest photon index of $3.15\pm0.03$.
    Furthermore, we calculate the disk flux contribution 
    in the total spectrum for all observations. It is found that
    Obs G spectrum has the largest disk flux contribution of $\sim45\%$
    in net flux in $0.1-100.0$ keV, which  
    indicates dominance of the disk and major contribution of soft 
    thermal component in the spectrum. These 
    findings evidently suggest that Cygnus X-1 
    is in extremely soft spectral 
    state. Also, during this observation, we notice that the 
    {\it MAXI} hardness ratio drops to a very low value of $\sim 0.229$ 
    (\autoref{fig:maxi_plot_figure}), which further corroborates 
    an extreme soft nature of the source. Studying
    the observed variability of {\it MAXI} hardness ratio of last 
    $10$ years we infer that Cygnus X-1 possibly attains an `extreme' soft 
    state that has ever been recorded. It further motivates us to 
    analyse and study timing properties with {\it AstroSat} observations.
	
    \section{Timing analysis and results}
    \label{sec:timing}

	\begin{figure*}
		\begin{center}
		\includegraphics[scale=0.52]{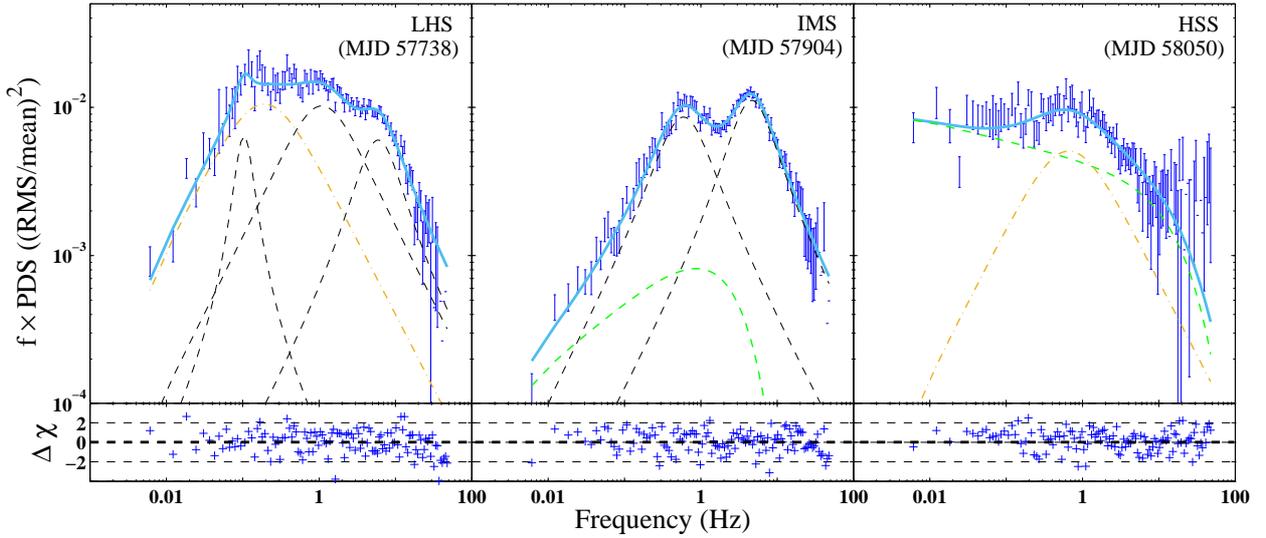}
	    \caption{Model fitted PDS (solid cyan line) in frequency $0.006- 50.0$ Hz of 
	    low-hard state (left), 
	    intermediate state (middle) and high-soft state (right) 
	    of Cygnus X-1 for observation B, D and G unfolded against the models. The fits
	    are performed for the energy range from $3.0-50.0$ keV using {\it LAXPC} data.
	    The fit model includes  multiple {\tt Lorentzians} (black and yellow dashed line)
	    for narrow or board profiles plus cut-off power-law component (dashed green line).
	    Zero centered  {\tt Lorentzians} represent broadband limited noise (BLN) shown 
	    in yellow dashed line. Bottom panels show the corresponding residuals in units of 
	    $\sigma$. See text for details.}
	    \label{fig:pds_figure}
	    \end{center}
	\end{figure*}
	
    \begin{table*}
    \caption{The best-fit parameters results from fitting models to PDS.
    		From left to right, are
    		(1) name of observation;
	   		(2) width of zero centered {\tt Lorentzians} representing BLN;
		    (3)-(8) frequencies and corresponding widths of {\tt Lorentzians}; 
		    (9) index of cut-off power-law;
		    (10) fractional RMS of BLN component in percentage;
		    (11) total fractional RMS;
		    (12) fractional flux variability;
		    (13) $\chi_{red}^{2}\,(\chi^{2}$/dof; dof $\equiv$ degree of freedom) resulting from fit.
		    See text for details.} 
    \label{tab:pds_fit_table} 
		\scalebox{0.81}{
		\begin{tabular}{lcccccccccccc} 
			\hline
			\hline
			Obs & $\sigma_{1}$ & $\nu_{2}$ & $\sigma_{2}$ 
			& $\nu_{3}$ & $\sigma_{3}$ & $\nu_{4}$ & $\sigma_{4}$ 
			&  $\alpha$ & RMS$_{BLN}$	&   RMS$_{total}$ &F$_{var}$& $\chi_{red}^{2}$($\chi^{2}$/dof)\\
			&(FWHM)&(Hz)&(FWHM)&(Hz)&(FWHM)&(Hz)&(FWHM)&&(\%)&(\%)&(\%)&\\
			\hline
			Obs A      & $0.32_{-0.03}^{+0.04}$ & $0.60_{-0.17}^{+0.13}$ & $2.04\pm0.21$ 
			& $4.44\pm0.36$ & $7.27_{-0.33}^{+0.31}$ & $0.07\pm0.01$ & $0.010_{-0.007}^{+0.008}$
			&-                          & $20.64$ & $28.16$ & $37.7$& $1.29(144.09/111)$\\
			Obs B      & $0.37_{-0.03}^{+0.07}$ & $0.32_{-0.22}^{+0.21}$ & $2.03_{-0.14}^{+0.21}$
			& $4.54_{-0.42}^{+0.36}$ & $7.70_{-0.37}^{+0.36}$ & $0.1$(fixed) & $0.06_{-0.02}^{+0.03}$
			&-                          & $17.87$ & $27.60$ &$33.5$ & $1.66(184.63/111)$\\
			Obs C      &     -                     & $0.42_{-0.05}^{+0.06}$ & $1.45_{-0.10}^{+0.07}$
			& $4.6\pm0.08$ & $5.80_{-0.15}^{+0.16}$ & -                       & -
			& $0.68_{-0.21}^{+0.22}$ &    -  & $20.11$ & $38.2$& $1.54(176.19/114)$\\
			Obs D      &     -                     & $0.38\pm0.03$ & $0.90\pm0.06$
			& $3.44\pm0.09$ & $6.21\pm0.14$ & -                       & -
			& $0.56_{-0.08}^{+0.11}$ &    -  & $20.47$ &$33.8$& $1.28(146.40/114)$\\
			Obs E      &-&-&-
			&-&-&-&-
			& $0.97\pm0.01$ &    -  & $25.94$ &$28.3$ & $1.58(188.30/119)$\\
			Obs F      & $0.39_{-0.05}^{+0.06}$ &-&-
			&-&-&-&-
			& $0.96\pm0.02$ & $13.47$ & $28.43$ &$27.7$ & $1.72(201.70/117)$\\
			{\bf Obs G}      & $1.32\pm0.13$ &-&-
			&-&-&-&-
			& $1.12\pm0.04$ & $12.59$ & $24.99$ &$20.1$ & $1.32(155.07/117)$\\
			Obs H      & $0.54_{-0.06}^{+0.07}$ &-&-
			&-&-&-&-
			& $0.98\pm0.01$ & $10.66$ & $27.94$ &$21.1$& $1.47(172.45/117)$\\
			%Obs I &&&&&&&&&&&&\\
			\hline
			\end{tabular}}
	\end{table*}
	
	We track the evolution of temporal features of Cygnus X-1 strictly simultaneous
	to spectral observations. In order to study temporal features of source in different 
	states, lightcurves of $5$ ms resolution from {\it LAXPC20} data 
	sets in the energy range from $3.0 - 50.0$ keV  are extracted.
	Thereafter  each lightcurve is divided in multiple intervals of $2^{15}$ timebins 
	(equivalent to multiple $160$ sec long lightcurve segments) and then PDS
	is generated for each interval. The final PDS for each lightcurve is generated 
	after averaging these individual PDS over entire observation. The averaged PDS is 
	binned by a geometrical factor of $1.05$ in frequency space. Normalization for each PDS 
	is similar to that of \citet{1990A&A...227L..33B, 1992ApJ...391L..21M}
	where PDS is given in units of the squared fractional rms variability per
	frequency interval. Further, we represent the PDS in units of frequency times 
	power ({\it f $\times$ P$_{f}$}) versus frequency \citep{1997A&A...322..857B}.
	We consider frequency range upto $50$ Hz for modelling as beyond this 
	in higher frequencies data is not reliable and may be affected by dead time correction.
	
	We start with modelling of power density spectra of LHS (Obs A), with three {\tt Lorentzians}
	since previous investigators \citep{2003A&A...407.1039P, 
	2005A&A...438..999A, 2017ApJ...835..195M} have shown that PDS of Cygnus X-1 in 
	hard state can be well modelled with multiple {\tt Lorentzian} profiles. One of 
	the {\tt Lorentzians} center is fixed to zero to represent broad band limited noise (BLN). 
	This model gives a $\chi_{red}^{2}$ of $154.85/114 = 1.36$. The residuals 
	suggest requirement of a narrow {\tt Lorentzian} feature below $0.1$ Hz.
	Hence an additional {\tt Lorentzian} component is added to the model.
	Fit results in an improved $\chi_{red}^{2}$ value 
	of $144.09/111 = 1.29$. We consider this model as the best fit for LHS and 
	similarly model other PDS of LHS obtained for Obs B. Fit results are 
	shown in \autoref{fig:pds_figure} (left) and best fit parameters are 
	given in \autoref{tab:pds_fit_table}. We calculate fractional 
	root mean square (RMS) variability by integrating model
	over frequency range of $0.006-50.0$ Hz.
	An RMS value of $\sim 28\%$ is noted in LHS, of which
	$\sim 20\%$ contributes to BLN component. Further, PDS of IMS from Obs C and D are fitted 
	with two {\tt Lorentzian} as power spectra show only two broad features. A $\chi_{red}^{2}$ 
	of $215.57/116 = 1.85$ results from the fit for Obs C. Residuals below $0.2$ 
	Hz indicate	a need of power-law to further improve the fit. Hence a cut-off 
	power-law is incorporated in existing model which gives a 
	$\chi_{red}^{2}$ of $176.19/113 = 1.54$. Similar fit procedure is
	applied to PDS from Obs D. We calculate an RMS value of $\sim20\%$
	in PDS for these observations, which is $\sim8\%$ lesser than that of LHS.
	
	Moving onto softer spectral states observations, we notice that PDS 
	are steep in nature. Hence PDS, obtained 
	from Obs E to Obs H, are modelled with a cut-off power-law. We notice fit for Obs E
	results in $\chi_{red}^{2}$ of $188.3/119 = 1.58$. The power-law index value
	is found to be $0.97\pm0.01$. It is observed that RMS value increases
	to $\sim 26\%$ when compared that of Obs C \& D.
	Obs F, G \& H when fitted with power-law, show residual suggesting 
	presence of BLN component. Addition of BLN component to power-law
	improves the fit significantly for Obs F \& H and fits result in 
	$\chi_{red}^{2}$ values of $1.72$ \& $1.47$, respectively.
	This improvement is moderate for
	Obs G with a $\chi_{red}^{2}$ value of $1.32$. 
	Overall fit for Obs G is shown in
	right panel of \autoref{fig:pds_figure}.
	We notice fits give power-law index close to unity 
	($\propto$ {\it $f ^{-1}$}). 
	
	Moreover, we observe during Obs G when spectral nature 
	becomes the softest, cut-off power-law index is further steeper with an index of 
	$1.12\pm0.07$, when compared to that of other observations (Obs F and H) with soft spectral 
	nature. In terms of RMS values Obs F \& H have
	roughly same values of $\sim 28\%$, while Obs G shows a lesser
	value of $\sim 25\%$. Finally, we calculate the flux variability 
	(\(F_{var} = \sqrt{\frac{S^2 - \bar{\sigma}_{err}^{2}}{\bar{x}}}\), 
	where S$^2$, \(\bar{\sigma}_{err}^{2}\) and 
	\(\bar{x}\) are the variance, mean square error and arithmetic mean 
	of time series representing lightcurve) 
	for each observation following \citet{2003MNRAS.345.1271V}.
	The calculated F$_{var}$ for various states are presented
	in \autoref{tab:pds_fit_table}.
	
	\section{The continuum-fitting and the softest state of Cygnus X-1}
	\label{sec:softest_spec_analysis}
	
	\subsection{The Softest Nature of Cygnus X-1: {\it AstroSat} and Previous Observations}
	\label{sec:softest_comparison}	
	\begin{figure}		
		\includegraphics[width=\columnwidth]{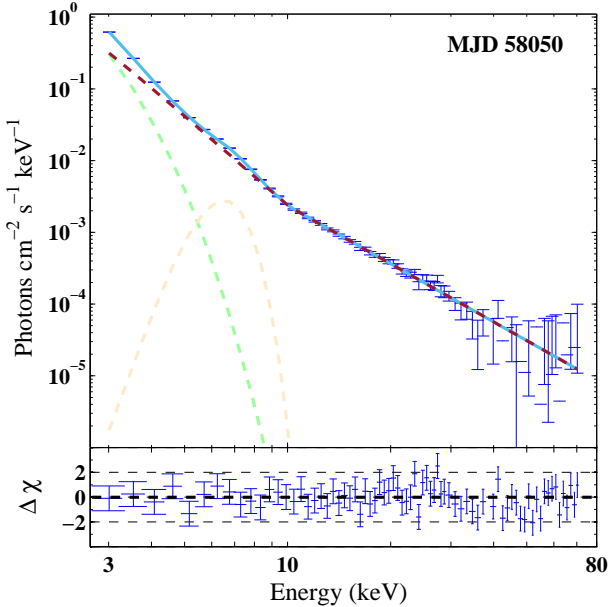}
	    \caption{The softest spectrum of Cygnus X-1, from observation on 
	    Oct $24,\,2017$ (MJD $58050$), fitted with an absorbed disk ({\tt diskbb})
	    plus broken power-law ({\tt bknpower}). Overall fit is shown with 
	    solid cyan line. The disk, broken power-law and 
	    Gaussian profile of Iron line is shown with green, brown 
	    and orange dashed line, respectively. 
	    The lower panel shows the fit residual in units of 
	    $\sigma$. The scheme prescribed by \citet{2006A&A...447..245W}
	    is followed to estimate the low energy spectral index 
	    $\Gamma_{1}$ ($\sim4.13$). See $\S$\ref{sec:softest_spec_analysis} 
	    for details.}
	    \label{fig:wilms_plot_figure}
	\end{figure}

	Cygnus X-1, although persistently remains in
	hard state but occasionally it transits to 
	extreme soft states. 
	The spectral analysis described in $\S$\ref{sec:spectral_analysis} 
	indicates that {\it AstroSat} 
	observation of Cygnus X-1 on Oct 24, 2017 (Obs G) is very distinctive in 
	terms of spectral softness and the spectral characteristics, 
	which exhibits an extremely soft nature. 
	In order to carry out a comparative study
	of this soft nature of the source with 
	previously reported as the softest state 
	by \citet{2017PASJ...69...36K} from {\it Suzaku} 
	observations on May 07, 2013, an in-depth analysis
	of the {\it AstroSat} data is performed. 
	
	Firstly, {\it MAXI}-HR for Oct $24,\,2017$ 
	(MJD $58050$) is obtained and it is compared with the same 
	for May $07,\,2013$ (MJD $56419$). 
	We find that {\it MAXI}-HR drops to $\sim0.229$ 
	during {\it AstroSat} observation, while the same 
	corresponding to {\it Suzaku} observations is at higher 
	value of $\sim0.238$ (see \autoref{fig:maxi_plot_figure}). 
	The lower {\it MAXI}-HR  value is an indicative of 
	more soft energy photon flux relative to that of in hard energy.
	Secondly, {\it LAXPC} lightcurve is also examined and we  
	obtain the hardness ratio (hereafter {\it LAXPC}-HR)
	of photon counts in energy range 
	$10.0-50.0$ keV to $3.0-10.0$ keV. 
	We choose these energy ranges as 
	it is noted in the spectral analysis  ($\S$\ref{sec:spectral_analysis})
	that disk component extends upto $\sim10$ keV whereas
	the photon flux in $10.0-50.0$ is attributed to Comptonisation component.
	Hence, the {\it LAXPC}-HR reflects variation of the 
	relative contributions of disk component and Comptonised emission
	component in total spectrum with time.
	We notice a time interval where the {\it LAXPC}-HR remains
	about a minimum value of $\sim0.27$.
	Subsequently, {\it SXT} and {\it LAXPC} spectra from Obs G data
	are extracted for this time interval only. 
	This extracted spectrum is further used to compare
	the spectral softness of the source with that of 
	described in \citet{2017PASJ...69...36K}. 
	We follow the same method 
	adopted by \citet{2017PASJ...69...36K}
	and fit an absorbed disk plus broken 
	power-law model over {\it LAXPC} spectrum ($3.0-70.0$ keV).
	The method, defining the spectral nature 
	in Cygnus X-1, is prescribed by \citet{2006A&A...447..245W}
	and \citet{2014A&A...565A...1G}, which is based 
	on low energy index of broken power-law.
	The low energy photon index $\Gamma_{1}$ suggests
	the disk dominance over high energy tail.		
	We estimate the low energy photon index 
	$\Gamma_{1} = 4.13^{+0.06}_{-0.08}$, 
	which is even higher than that of the softest 
	state reported by \citet{2017PASJ...69...36K}
	as $\Gamma_{1} \sim$ 4.0.
	The model fitted LAXPC spectrum is shown 
	in \autoref{fig:wilms_plot_figure}.
	
	Combining the results from these two methods,
	it is evident that Cygnus X-1 during {\it AstroSat}
	observation on MJD $58050$ (Obs G) exhibits a minimum value 
	of {\it MAXI}-HR and a steeper of low energy photon index 
	($\Gamma_{1}$) than those of during {\it Suzaku} observations.
	The two findings corroborate the highest contribution 
	of soft disk component in total spectrum ever observed
	from the source.
	
	\subsection{Constraining the Spin of Cygnus X-1}
	
	\label{sec:spin_kerr_fit}
	\begin{figure}
		\includegraphics[width=\columnwidth]{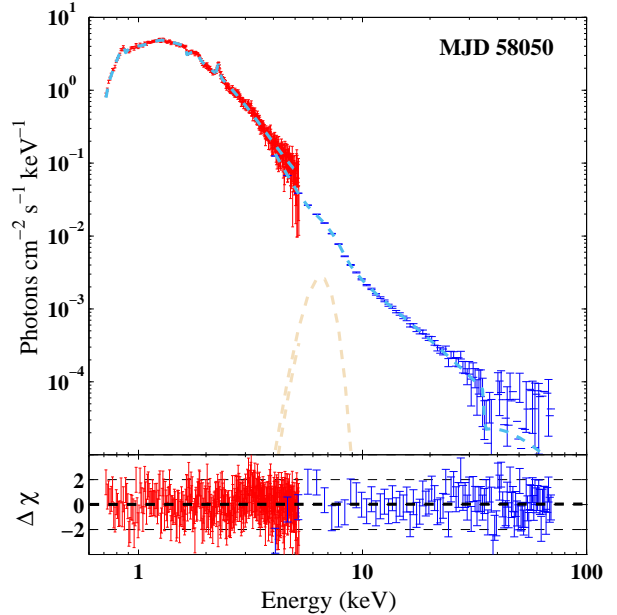}
		
	    \caption{Spectrum of Cygnus X-1, observed on Oct $24,\,2017$
	    (MJD $58050$), fitted with relativistic model. The model 
	    comprises of an absorbed {\it kerrbb} convoluted with {\it simpl}. 
	    Cyan dashed line shows the total model.
	    Gaussian profile of Iron 
	    line is shown with orange dashed line.
	    Bottom panels show the corresponding residuals 
	    in units of $\sigma$. See $\S$\ref{sec:spin_kerr_fit} for details.}
	    \label{fig:kerbb_spec}
	\end{figure}
	
	\begin{table*}
			\captionsetup{justification=centering, margin =2cm}
			\caption{Spectral fit results with relativistic model 
			{\tt tbabs*(simpl$\otimes$kerrbb+gaussian)}. Parameters are (top to bottom)
			(1) hydrogen column density in units of $\times 10^{22}$ atoms cm$^{-2}$;
		    (2) photon power-law index; 
		    (3) scattering fraction in percentage;
		    (4) spin parameter;
		    (5) mass accretion rate in units of $\times 10^{18}$ gm s$^{-1}$; 
		    (6) mass of black hole in units of M$_{\sun}$;
		    (7) distance to Cygnus X-1 in units of kpc;
		    (8) inclination of binary plane in degrees;
		    (9) spectral hardening factor.
		    Orbit number provided for Obs G and I are the spacecraft
		    orbit numbers.}
			\label{tab:kerr_fit_param_table}
			\begin{tabular}{lcccc} 

		    \hline
			\hline
			Components 		&   Parameter & \multicolumn{2}{c}{Obs G} & Obs I \\
					
							&& Orbit 11212 & Orbit 11215 & Orbit 15548\\  
			\hline
			{\it tbabs} & N$_{H}$		    				&$0.67\pm0.01$&$0.66\pm{0.02}$
															&$0.68_{-0.01}^{+0.02}$\\
			{\it simpl} & $\Gamma$					  		&$2.98\pm{0.02}$&$2.21\pm{0.03}$
															&$2.61\pm{0.01}$\\
			& {\it f$_{sc}$}		  						&$10.56_{-0.31}^{+0.22}$&$5.56\pm{0.01}$
															&$11.08_{-0.21}^{+0.32}$\\
			{\it kerrbb} & a$_{*}$							&$0.9998_{-0.0015}^{+0.0000}$
			       											&$0.9999_{-0.0121}^{+0.0000}$
			       											&$0.9998_{-0.0155}^{+0.0000}$\\
			    		& $\dot{\rm{M}}$							&$0.14_{-0.01}^{+0.02}$
			    											&$0.12\pm{0.01}$
			    											&$0.11\pm{0.02}$\\
			      	 	& M$\rm{_{BH}}$			 			  	&$21.2$ (fixed)&$21.2$ (fixed)
			      	 										&$21.2$ (fixed)\\
			       		& D$\rm{_{BH}}$							&$2.22$ (fixed)&$2.22$ (fixed)
			       											&$2.22$ (fixed)\\ 
			       		& $i$                              	&$27.5$ (fixed)&$27.5$ (fixed)
			       											&$27.5$ (fixed)\\
			       		& {\it f} 							&$1.73_{-0.01}^{+0.01}$
			       											&$1.74_{-0.02}^{+0.01}$
			       											&$1.82_{-0.02}^{+0.01}$\\
			\hline
			            &$\chi_{red}^{2}$ ($\chi^{2}$/dof)   &$1.32\,(712.8/540)$&$1.31\,(719.6/546)$
			            									&$1.42\,(732.71/516)$\\
			                     
			\hline                     
		\end{tabular}
	\end{table*}
	
	In previous sections ($\S$\ref{sec:softest_state}
	and $\S$\ref{sec:softest_comparison}),
	it is shown that the source exhibits an extreme soft state 
	during {\it AstroSat} observations (Obs G).
	Further, the continuum-fitting method is employed and spectrum of HSS 
	is modelled with relativistic model to estimate	the spin. 
	For relativistic modeling of spectrum, we 
	follow \citet{2014ApJ...790...29G} and \citet{2017PASJ...69...36K}
	and use {\tt constant * tbabs(simpl $\otimes$ kerrbb + gaussian)} 
	model to fit the spectrum. {\tt simpl} is a 
	Comptonising convolution model, which produces power-law distribution
	in higher energies 
	using soft seed photons from accretion disk independently irrespective of any 
	shape or location of Comptonising medium \citep{2009PASP..121.1279S}. 
	The {\tt simpl} estimates two parameters, 
	{\it f$_{sc}$} and $\Gamma$. {\it f$_{sc}$} gives the fraction of seed
	photons up-scattered to power-law tail while $\Gamma$ estimates the steepness of 
	this tail. {\tt kerrbb} is a relativistic thin accretion disk model 
	which includes self-irradiation and limb-darkening effects 
	\citep{2005ApJS..157..335L}. We switch off limb-darkening and apply zero torque 
	condition at the inner boundary of the disk. Moreover, model is provided with 
	updated measurements of distance to source (D), mass of the black hole (M$\rm{_{BH}}$) in 
	the binary system and inclination of the binary plane ({\it i}) 
	as mentioned in $\S$\ref{sec:intro}
	and thus the normalization is fixed to unity. The other two parameters, 
	namely accretion rate ($\dot{\rm{M}}$) and spectral hardening factor
	({\it f}) are allowed to vary freely. The {\tt gaussian} component
	is added to model the Iron line profile. Further, in order 
	to achieve acceptable fit results, we correct for instrument 
	features at $1.8$ and $2.24$ keV arising from SXT. Also, a broad absorption dip 
	present at $\sim 8$ keV	is modelled with {\tt smedge}. The overall fit results in 
	$\chi_{red}^{2}$ of $712.8/540 = 1.32$. The model fitted spectrum is
	shown in \autoref{fig:kerbb_spec}. $\dot{\rm{M}}$ is estimated as 
	$\sim0.14\times10^{18}$ g s$^{-1}$. Spectral 
	hardening factor ({\it f}) is determined as $1.73\pm0.01$, which is in good 
	agreement with values suggested for stellar mass black holes \citep{1995ApJ...445..780S}. 
	The scattering fraction 
	{\it f$_{sc}$} comes to be $10.56^{+0.22}_{-0.31}\%$, which implies that
	the data is of good quality and it well satisfies the criterion for continuum-fitting method 
	\citep{2009ApJ...701L..83S}.
	The fit gives the power-law index ($\Gamma$) of $2.98\pm0.02$, which is comparable with
	the estimate of $\Gamma = 2.93^{+0.11}_{-0.05}$ by \citet{2017PASJ...69...36K}. 
	This again indicates that during {\it AstroSat} observation Cygnus X-1 shows 
	the weakest Comptonisation tail.  The best fit gives an extreme value of the spin  
	parameter as a$_{\ast}=0.9998_{-0.0015}^{+0.0000}$,  which is very close to the
	hard limit of {\tt kerrbb} model. In order to estimate the errors on spin parameter
	we perform Monte Carlo (MC) simulations,  which is discussed later in the section.
	Additionally, the similar fit procedure is applied
	to two more observations (Obs G, Orbit no. $11215$ and Obs I) where spectra are 
	suitable for spin estimations. 
	Although these spectra have less steeper $\Gamma$ than that of Obs G (Orbit no. 11212), 
	however, the fits give acceptable {\it f$_{sc}$} values, which further made them 
	reliable for spin estimation. The spectral fit parameters are provided in 
	\autoref{tab:kerr_fit_param_table}.

	\begin{figure*}
	 	\begin{center}
	 	\hskip -5 mm	
		\includegraphics[width=1.02\textwidth]{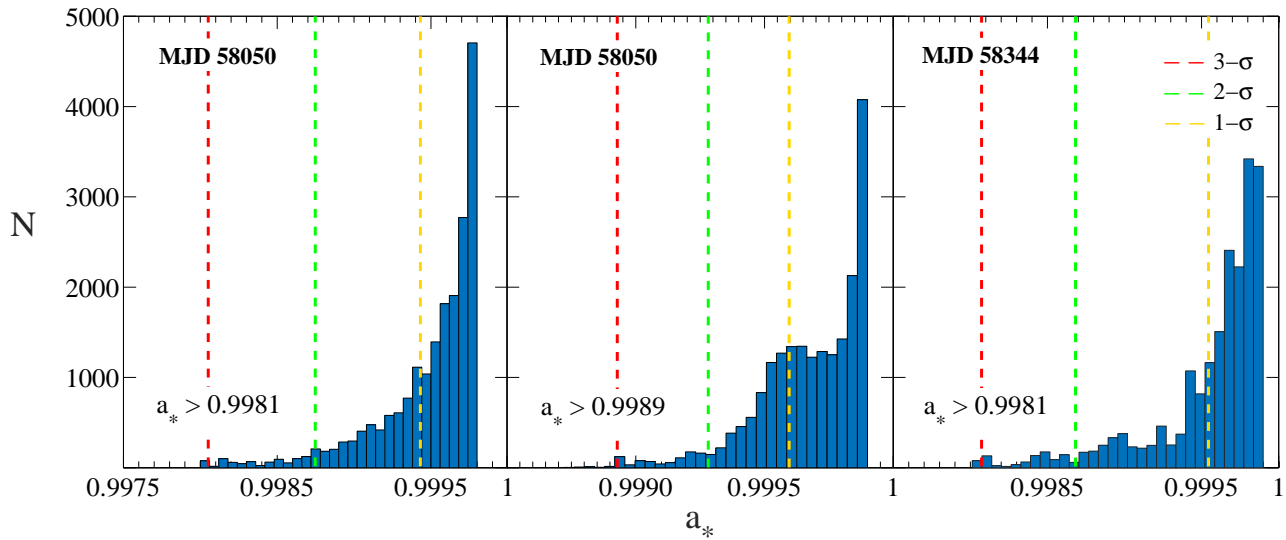}
	    \caption{The histograms of spin parameter (a$_{*}$) resulting 
	    from Monte Carlo simulations with 20000 sets of parameters, containing 
	    mass of the blackhole (M$\rm{_{BH}}$), inclination of the binary ({\it i})
	    and distance to the source (D), per
	    spectrum. The three vertical dashed lines, from left to right 
	    in each panel, mark the error confidence of $3\sigma\,(99.7\%$, red), 
	    $2\sigma\,(95.4\%$, green) and $1\sigma\,(68.3\%$, yellow). The 
	    corresponding $3\sigma$ confidence lower limits of a$_{*}$ is also
	    given in each panel. The left, middle and right histogram panels
	    are obtained from observation G (orbit no. 11212), G (orbit no, 11215) and
	    I, respectively. See text for details.}
	    \label{fig:spin_histogram}
	    \end{center}
	\end{figure*}

	Further, to estimate the uncertainties in spin parameter values,
	obtained from each spectra, we perform MC simulations and 
	generate 20000 sets of binary system parameters (M$\rm{_{BH}}$, {\it i} and D),
	as previous spin estimation works suggest that uncertainties in
	these three input parameters dominate over the errors resulting
	from {\tt kerrbb} model \citep{2011ApJ...742...85G,
	2014SSRv..183..295M, 2021ApJ...908..117Z}.
	The sets are generated assuming independent Gaussian-distributions
	for each parameter.
	Subsequently, we calculate the look-up table of hardening factor (f),
	corresponding to each parameter set and then fit the spectra to determine spin parameters.
	This gives 20000 values of spin parameters per spectrum. We repeat
	this procedure on all three observations and finally, obtain
	respective histograms of a$_{*}$, which are shown in \autoref{fig:spin_histogram}.
	We consider the lowest spin parameter value of all the three observations,
    for the final estimation of spin as
	a$_{*}> 0.9981 $ at  $3\sigma$ confidence level. We also calculate
	the spin parameter with a summed up histogram of all the observations, which
	also results into the similar value with $3\sigma$ confidence. The lower limits
	for $2\sigma$ and $1\sigma$ are $0.9988$ and $0.9995$, respectively. Similar 
	lower limits for individual spectrum are marked in \autoref{fig:spin_histogram}.

	\section{Discussion and Conclusion}
	\label{sec:discussion}
	{\it AstroSat} made several co-ordinated observations 
	on Cygnus X-1, starting from $2016$ to $2020$. Initially in 
	the year $2016$ the source is in low-hard state.
	In March, $2017$, the source starts to show signature of transition from the 
	persistent low-hard state. The simultaneous and continuous monitoring 
	by {\it MAXI} also exhibits a clear transition 
	with a decreasing spectral hardness ratio value 
	from its mean value of $\sim0.8$ 
	(see \autoref{fig:maxi_plot_figure}). The hardness ratio
	consistently keeps decreasing till Oct, $2017$, while the source 
	passes through intermediate state and attains a high-soft state.
	Cygnus X-1, in a period of about $7$ months starting 
	from March, $2017$ till Sept, 2017, completely transits to 
	a high-soft state. The source remains in soft state for about 
	two months. We observe that the hardness 
	ratio of the source decreases to the lowest value of $\sim 0.229$.
	From Nov, $2017$ onwards, it is observed
	that there is increase in hardness ratio which marks the transition back to 
	the low-hard state. Subsequently, the source shows flickering sequence of 
	transitions from hard to soft states and vice versa (see \autoref{fig:maxi_plot_figure}).
	{\it AstroSat} observations sample the transition of Cygnus X-1 
	in year $2017$ from LHS
	to HSS via IMS (see \autoref{fig:maxi_plot_figure} and \autoref{tab:obs_table}).
	We make use of data from {\it SXT} and {\it LAXPC} 
	X-ray instruments on-board {\it AstroSat} and study the evolution 
	of spectral and temporal properties of the source during the transition
	including hard and extreme soft states.
	
\begin{figure}
		\includegraphics[width=\columnwidth]{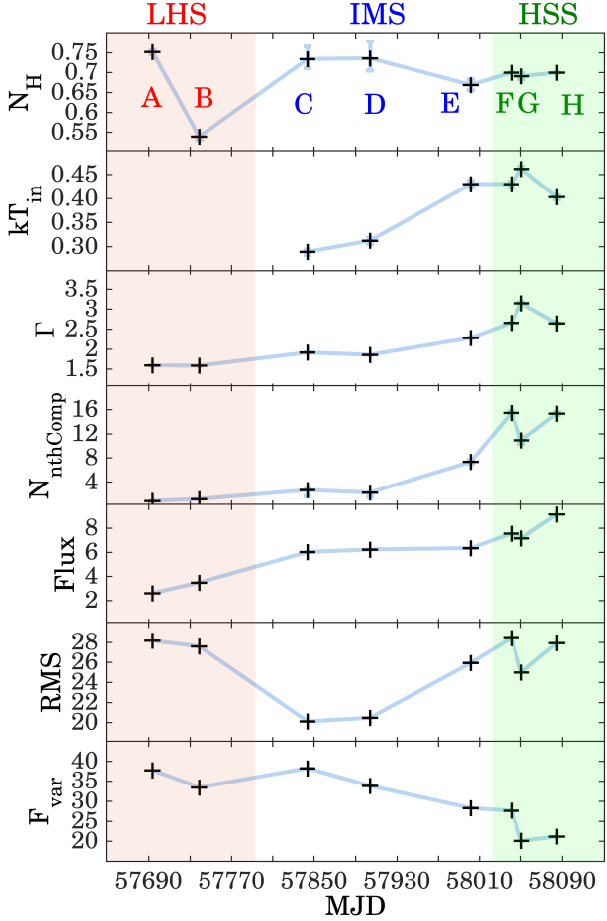}
	    \caption{The evolution of spectral and timing
	    parameters of Cygnus X-1, during transition
	    from LHS to HSS via IMS.
	    The parameters are (top to bottom) hydrogen column
	    density (N$_{H}$) in units of $\times 10^{22}$ atoms cm$^{-2}$,
	    inner disk temperature (kT$_{in}$) in units of keV, photon index ($\Gamma$),
	    normalization constant of {\tt nthComp}, unabsorbed flux, calculated 
	    in $0.1 - 100.0$ keV energy range, in units 
		of $\times 10^{-8}$ ergs cm$^{-2}$ s$^{-1}$,
		total fractional RMS of PDS in percent,
		fractional flux variability in percentage (see text for details).
		Observations from A to H are marked and shown in
		top panel. Red shaded region represents LHS,
		blue is IMS and green shows HSS.
		For Obs A \& B inner disk temperature is not shown
		as the disk component is not required to model 
		the LHS	spectra.} 
	    \label{fig:param_variation}
	\end{figure}

	{\it SXT} \& {\it LAXPC} combined spectra are modelled in 
	$0.5$--$70.0$ $\rm{keV}$ energy range. Spectra from LHS of Cygnus X-1 
	are modelled with absorbed 
	non-thermal Comptonisation model 
	({\tt nthComp}) with out any signature of thermal component 
	in this state. Spectral indices ($\Gamma$) of $1.59\pm{0.01}$ 
	are found to be in good agreement with usual LHS spectral 
	index of Cygnus X-1
	\citep{2014A&A...565A...1G} as well as with other BH-XRBs
	(GX $339-4$ \citet{2012A&A...542A..56N}, IGR $17091-3624$ \citet{2018Ap&SS.363..189D}, 
	4U 1630-472 \citet{2020MNRAS.497.1197B}).
	In subsequent IMS,
	we find that spectral indices ($\Gamma)$ are steeper
	with values ranging in $1.86-2.28$.
	Along with that there is an indication of thermal disk 
	emission component ({\tt diskbb}) with inner disk temperature
	(kT$\rm{_{in}}$) in range of $0.29-0.43$ keV.	
	Further, during IMS, we note that with increase of spectral index,
	the inner disk temperature increases as well while the normalisation of disk 
	component decreases. This 
	could suggest that the disk approaches towards the central object. 
	Subsequent observations in Oct, 2017 show 
	that source is in HSS and the spectral indices are even
	steeper with values lying in $2.5-3.15$. During HSS, from 
	Obs G on Oct $24,\,2017$ (MJD $58050$), 
	we find that spectral index of source is the steepest with $\Gamma=3.15\pm0.03$
	along with an inner disk temperature (kT$\rm{_{in}}$) of $0.46\pm0.01$ keV.
	Similar steep spectral nature and inner disk temperature, 
	during HSS of other BH-XRBs have been reported previously. For 
	example, findings from the HSS of GX $339-4$ by \citet{2012A&A...542A..56N}  
	show an inner disk temperature and photon index of $\sim0.7$ keV and 
	$\sim3.5$ respectively. \citet{2018Ap&SS.363..189D} has shown 
	for IGR $17091-3624$ that the inner disk temperature reaches
	upto $1.3$ keV in HSS.
	Further, we observe that the inner disk temperature increases consistently with
	increase in spectral index from LHS to HSS, which  is shown in \autoref{fig:param_variation}.
	Along with this, we note that the fractional disk flux contributions
	in net spectra vary in $33\%-45\%$ during IMS to HSS.
	The maximum fractional disk contribution of $\sim\!45\%$
	is observed in Obs G of HSS.
	It is to be noted that the model with combination 
    of {\tt diskbb} and {\tt nthComp} results in 
    slightly higher values of $\chi_{red}^{2}$ ($>\!1.5$), indicating
    further investigation to improve modelling of the HSS spectra.
    Although, for uniformity and comparative study of spectral parameters
    we use the same model for all the states.
	Further, based on the inner disk temperature (kT$_{in}$) 
    and photon index ($\Gamma$)	we classify the observed spectra
    into three states, namely, LHS, IMS and HSS (see $\S$\ref{sec:softest_state}
    and \autoref{fig:param_variation}).
    We also estimate 
    the unabsorbed flux $0.1-100.0$ keV energy range for all states 
    (see \autoref{fig:param_variation}) and compare it
    with Eddington luminosity. We notice that luminosity of the source varies 
	from $0.3\%$ to $\sim 1.4\%$ of Eddington luminosity
	(L$_{edd}$\footnote{\label{note:L_edd}For Cygnus X-1 L$_{edd}\approx  2.8
    \times10^{39}$ egs s$^{-1}$})
    of Cygnus X-1. The luminosities during different states
    of the source are in good agreements with previous findings by
    \citet{2011ApJ...742...85G, 2013PASJ...65...80Y, 2014ApJ...790...29G, 
    2014ApJ...780...78T, 2016ApJ...826...87W, 
    2017MNRAS.472.4220B, 2018ApJ...855....3T}.
    
	Also, we examine the power density spectra (PDS) of Cygnus X-1 simultaneous 
	to spectral analysis. PDS of LHS \& IMS show multiple broad {\tt lorentzians}
	features along with band limited noise.
	In soft states of the source, we observe that broad   
	features are absent and PDS show a steep 
	cut-off power-law with index of $\sim\!1$ 
	and again a band limited noise is present.
	The temporal features of Cygnus X-1 are in good agreement with
	previous studies by \citep{2003A&A...407.1039P,2005A&A...438..999A,
	2017ApJ...835..195M}. Non-detection of QPO like
	features in HSS of Cygnus X-1 is quite common
	as the featureless PDS are observed in other BH-XRBs
	too (see \citet{2014AdSpR..54.1678R, 2018Ap&SS.363...90N} for XTE J$1859+226$,
	\citet{2016MNRAS.460.4403R} for V$404$ Cyg, \citet{2005A&A...440..207B,
    2012A&A...542A..56N} for GX $339-4$).
    
	Furthermore, we highlight the results of Obs G of HSS as 
	it's PDS shows the steepest power-law index with a value 
	of $1.12\pm0.04$. Also the band limited noise show the 
	largest FWHM of $1.32$ Hz. 
	We also calculate the flux variability (F$_{var}$) in lightcurves 
	during all the observations and note an 
	anti-correlation with corresponding RMS calculated from 
	PDS modelling (see \autoref{fig:param_variation}). 
	The variation of RMS
	is also noticed with spectral index ($\Gamma$),
	which agrees to that of \citet{2014A&A...565A...1G}.
	During HSS, an RMS value of $\sim\!25\%$ is observed 
	which is comparable to that of in LHS.
	Display of such high RMS is a typical characteristics 
	of Cygnus X-1 \citep{2014A&A...565A...1G} and unlike
	the general nature of BH-XRBs \citep{2010LNP...794...53B}, 
	RMS does not drop to very low values.
	For example, GX $339-4$ and IGR $17091-3624$
	exhibit a low RMS of $\sim3\%$ in HSS 
	\citep{2012A&A...542A..56N,2018Ap&SS.363..189D}.
	Recently, \citet{2020arXiv201003782S} show
	in case of HSS of GRS $1915+105$, a total RMS 
	less than $2\%$.
	
	Moreover, we find that the high-soft state of Cygnus X-1 
	is unprecedented and {\it AstroSat} on Oct $24,\,2017$ (MJD $58050$)
	observed the source in an extremely soft state.	
	We compare the soft nature of source during {\it AstroSat} era 
	with previously reported the softest state of Cygnus X-1 
	by \citet{2017PASJ...69...36K} with {\it Suzaku} observations
	on May $07,\,2013$ (MJD $56419$). We note the following
	two criteria for the comparison.
	Firstly, the hardness ratio obtained from {\it MAXI} on MJD 58050
	with the lower value of $\sim0.229$ indicates the {\it AstroSat}
	observations are softer than that of during {\it Suzaku} observations.
	Secondly, we adopt similar method of modelling spectra with
	broken power law in range
	$3.0-70.0$ keV as adopted by \citet{2017PASJ...69...36K}.
	We find that the low energy photon index ($\Gamma_{1}$) of 
	broken power-law is 
	even steeper with a value of $4.13^{+0.06}_{-0.08}$ than that 
	of estimated by \citet{2017PASJ...69...36K} as $\sim\!4.0$, 
	which confirms the largest contribution of disk component
	in net spectrum ever recorded.
	Hence, in this work, we confirm the softest state of 
	Cygnus X-1 ever observed. It is to be noted that
	Cygnus X-1 occasionally enters in a high-soft state
	and it has never reached 'canonical' soft state.

	Further, we make use of the high quality data from {\it AstroSat}
	observations of the softest state of the source  
	to constrain the spin of black hole as the spectra
	are dominated with disk component and are suitable 
	for continuum-fitting (CF) method. We fit the
	spectra with {\tt kerrbb} convoluted with {\tt simpl} 
	in order to constrain the spin parameter (a$_{\ast}$) 
	of the black hole. Furthermore, we perform MC simulations
	to estimate the uncertainty in a$_{\ast}$, combining the
	errors in the binary system parameters, M$_{\rm BH}$, {\it i} and D.
	We report the spin parameter of Cygnus X-1 as  
	a$_{*}> 0.9981 $ at  $3\sigma\,(99.7\%)$ confidence level,
	which suggests an extreme spin of the black hole in
	Cygnus X-1.
	
	We compare the spin measurement results with those obtained by
	\citet{2021ApJ...908..117Z} recently, as a$_{*}> 0.9985\,(3\sigma)$.
	The spin estimation in this work
 	is in good agreement with their results.
 	Also, \citet{2020JHEAp..27...53Z} applied CF method
	on large set of spectra obtained from {\it HXMT} and 
    estimated lower limit of the spin parameter as a$_{\ast}> 0.967\,(3\sigma)$
    using revised values of M$\rm{_{BH}}$, {\it i} and D.
	
	Previously,	\citet{2014ApJ...790...29G} employed CF method and 
 	estimated a$_{\ast}> 0.983$ at a confidence level of 
 	$3\sigma\,(99.7\%)$. \citet{2017PASJ...69...36K} 
 	determined a$_{\ast}=0.95\pm0.01$ with similar method using 
 	data from {\it Suzaku}. Spin estimation by 
 	\citet{2014ApJ...780...78T} with  with Fe-K$_{\alpha}$ 
 	line fitting method using {\it Suzaku } and {\it NuSTAR} 
 	data resulted as a$_{\ast}=0.9882\pm0.0009$
 	($90\%$ confidence level).

	Focusing on other parameters of best-fit, 
 	we state that the scattering fraction 
	({\it f$_{sc}$}) comes to be $10.56^{+0.22}_{-0.31}\%$.
	Furthermore, we observe from spectral fit that the 
	luminosity of the disk component is consistent 
	in HSS with L/L$_{Edd}$ $\sim 1.4\%$. These parameters
	satisfy the criterion for application of continuum-fitting 
	method to produce reliable results as stated in $\S$\ref{sec:intro}. 
	
	Moreover, these results of continuum-method are based on 
	Novikov-Thorne model. The model assumes a zero torque which 
	introduces uncertainties in spin estimation.
	These uncertainties are very small as a$_{\ast}$ tends to unity.  
	Also, the model originated errors are much smaller than the errors in 
	measurements of M$\rm{_{BH}}$, {\it i} and D,
	which is important in case of thin disks.
	Another possible error can be from pile-up in {\it SXT} detector during
    high flux soft states. In order to estimate pile-up, we fit only
    {\it LAXPC} spectra to find out any spectral hardening from pile-up.
    It is observed that photon index varies by less that $2\%$, which is
    expected due to channel difference in two detectors. Hence, we 
    conclude that spin estimations obtained are fairly reliable.
	    
    In order to summarize, in this paper, we present an in-depth 
    spectral and timing analysis results obtained from {\it AstroSat} 
    observations of Cygnus X-1 during transition from hard state
    to soft state via intermediate states. Moreover, we confirm the detection
    of the `softest' nature ever observed from the source with {\it AstroSat}.
    We make use of the high quality spectral data of the softest state 
    to constrain the spin parameter 
    of the black hole. Finally, 
    considering the present estimates of spin and mass, we conclude that
    Cygnus X-1 is a maximally rotating, massive black hole binary 
    source. This compact nature echoes to that of the enigmatic 
    galactic black hole source GRS $1915+105$, which represents 
    a standard template for BH-XRBs \citep{2020arXiv201003782S}.
    
    \section*{Acknowledgements}
	We are thankful to the anonymous referee for giving valuable suggestions
	to improve on spin estimation results. We also thank Sreehari 
	from Indian Institute of Astrophysics (IIA) for his help in modelling.
	This publication uses the data from the {\it AstroSat} mission of the Indian 
	Space Research Organisation (ISRO), archived at the Indian Space Science 
	Data Centre (ISSDC). This work has used the data from the Soft X-ray 
	Telescope ({\it SXT}) developed at TIFR, Mumbai, and the {\it SXT} POC at TIFR is 
	thanked for verifying and releasing the data and providing the necessary 
	software tools. This work has also used the data from the {\it LAXPC} Instruments 
	developed at TIFR, Mumbai, and the {\it LAXPC} POC at TIFR is thanked for verifying 
	and releasing the data. We thank the {\it AstroSat} Science Support Cell hosted by 
	IUCAA and TIFR for providing the {\it LaxpcSoft} software which we used for {\it LAXPC} 
	data analysis. Authors thank GH, SAG; 
	DD, PDMSA and Director, URSC for encouragement and continuous support to 
	carry out this research.
	
	\section*{Data availability statement}
	
	Data used for this work are available at 
	{\it AstroSat}-ISSDC website 
	(\url{http://astrobrowse.issdc.gov.in/astro_archive/archive}),
	and {\it MAXI} website
	(\url{http://maxi.riken.jp/top/index.html}).
	 
	%%%%%%%%%%%%%%%%%%%%%%%%%%%%%%%%%%%%%%%%%%%%%%%%%%
	
	%%%%%%%%%%%%%%%%%%%% REFERENCES %%%%%%%%%%%%%%%%%%
	
	% The best way to enter references is to use BibTeX:
	
	\bibliographystyle{mnras}
	\bibliography{references} % if your bibtex file is called example.bib

	\bsp	% typesetting comment
	\label{lastpage}
	\end{document}